\documentclass[review]{elsarticle}

\usepackage[T2A,T1]{fontenc}

\usepackage{hyperref}

\usepackage{xcolor}

\journal{JASTP}

\usepackage{array}
\usepackage{booktabs}

\biboptions{authoryear}

\begin{document}

\begin{frontmatter}

\title{Comparison of AATR and WTEC indices in the studies of the level of ionospheric disturbance}


\author{Berngardt O.I., Voeykov S.V., Perevalova N.P.}
\address{Institute of Solar-Terrestrial Physics SB RAS, 126a, Lermontova Str., Irkutsk, Russia 664033 }
\ead{berng@iszf.irk.ru,serg3108@iszf.irk.ru,pereval@iszf.irk.ru}

\begin{abstract}
A comparative statistical analysis of AATR and WTEC indices was conducted
based on data from the ISTP SB RAS GNSS receivers network. It is shown
that at high levels of ionospheric disturbance (for WTEC > 0.1 TECU),
the AATR index is proportional to the WTEC index with a factor of
$1.5min^{-1}$. At small levels of ionospheric disturbance (for WTEC
< 0.1 TECU), this proportionality is violated. It is shown that the
contribution of daily dynamics of the background ionosphere to the
AATR index is higher than to the WTEC index. This leads to a higher
sensitivity of the WTEC index to disturbances. This also leads to
violating the proportionality between WTEC and AATR indices at low
levels of ionospheric disturbance. It is shown that at high latitudes
the dynamics of the WTEC and AATR indices correlate significantly
with the level of geomagnetic disturbance Kp. At mid-latitudes, the
contribution of solar radiation variations (F10.7 index) and vertical
seismic variations exceeds the influence of Kp variations. The program
for calculating WTEC indices, used in the paper is put into open access. 
\end{abstract}


\end{frontmatter}


\section{Introduction }

The ionosphere and the ionospheric disturbances have a significant
impact on Global Navigation Satellite Systems (GNSS). Dense worldwide
networks of GNSS receivers and their high temporal resolution make
it possible to monitor the ionospheric conditions in real time using
Slant Total Electron Content (STEC) calculated along the line-of-sight
(LOS) between transmitting GNSS satellite and GNSS receiver. 

The methods of studying the ionosphere using GNSS receivers can be
divided into two large groups - techniques for using a single receiver,
and techniques for using a network of receivers (global or local).
In the case of using a network, it becomes possible to study
large-scale irregularities and their large-scale spatial variations, 
based on the production of total
electron content (TEC) maps: GIM \citep{Mannucci_1997,CDDIS_data} 
and TEC disturbance maps (TDM) \citep{Perevalova_2008,Afraimovich_2013}.
The first index (index of TEC perturbations) built based on GIM was 
proposed by \citep{Ho_1998}.
Later W and Wp indices have been developed
\citep{Gulyaeva_and_Stanislavswska_2008,GULYAEVA_2013,Stanislawska_2015}.
Using GIM allows one to distinguish the perturbed part in the world-wide distribution of TEC - nearly static and global-scale.
TDM allows one to to study wave-like disturbances associated with the passage of different irregularities of different scales \citep{Astafyeva_2018}. 
Joint analysis of network of receivers allows one to use radio tomographic methods to
determine the spatial characteristics of irregularities and to build
special indices based on them (HORT IPI \citep{Nesterov2017}) or use spatio-temporal combinations of their data, 
like DIX \citep{Jakowski_2012,Jakowski_2012b},
GIX and SIDX\citep{Jakowski_2019}, 
DIXSG\citep{Wilken_2018} and 
RIDX \citep{Stankov_2006} indices.

. 

Statistical parameters of mid- and
small-scale irregularities can be estimated using even a single GNSS receiver.
Using single receivers is useful when one have rarefied network, and minimizes the obtaining and data processing efforts.
Classical index for studying ionospheric irregularities is scintillation index S4, that is regularly measured by GNSS receivers\citep{Beniguel_2004}.
The S4 index is usually associated with small-scale irregularities,
comparable with the Fresnel radius, and leading to phase and amplitude
distortions of the received signal. 

To study larger ionospheric irregularities
that do not significantly distort the signal shape, one use the construction
of various integral indices based on variations of the measured signal
phase and group delays and characterized by STEC. STEC-based indices
include IROT \citep{Wanninger_1993}, ROTI\citep{Pi_1997}, fp and Fp
\citep{Mendillo_2000},
AATR \citep{sanz_2014,Juan_2018}, WTEC\citep{Voeykov_2016}.
The use of STEC-based indices looks more reasonable for analysis of mid-scale irregularities than S4. 

In the quiet mid-latitude ionosphere one of the sources of mid-scale
irregularities are acoustic waves, generated by different sources:
earthquakes \citep{Astafyeva_2018}, meteorites\citep{Berngardt_2015},
typhoones \citep{Polyakova_2013}, solar terminator \citep{Afraimovich_2009}, 
rocket launches \citep{Kakinami_2013,Zherebtsov_2019} and others.

Indices ROTI and AATR are traditional ones and can be used for
checking the effectiveness of other indices.

The ROTI index (Rate Of Change Of The TEC) has been used for monitoring
of mid-scale ionospheric disturbances for a long time \citep{Pi_1997}.
ROTI is
defined as the Root Mean Square error (RMS) of the Rate of Change
of the TEC (ROT) at a certain time interval (usually 5 min) at single
beam "satellite-receiver" and therefore
has local spatial character. 
A strong positive correlation between GNSS positioning error and ROTI
for receivers located above $64^{o}N$ \citep{Jacobsen_2014} allows
using it to estimate GNSS positioning errors at high-latitudes. 
High
values of ROTI index during severe geomagnetic storms in 2014-2015
\citep{Jacobsen_2016,Cherniak2018} allows using this index to monitor
the activity of mid-scale ionospheric irregularities at high- and
low-latitudes using single GNSS-receiver \citep{Pi_1997}. 

The AATR (Along Arc TEC Rate) index characterizes the average level
of ionospheric disturbances over wide spatial region, measured by
single GNSS-receiver \citep{sanz_2014,Juan_2018}.
Using AATR index is effective when one need to 
skip spatial dependence of ROTI.
In practice the AATR index
is defined as ROT averaged over a selected time interval (usually
over 1 hour) and averaged over all GNSS satellites visible at the
single receiver. The effective size of the area used for index estimation
is defined by relative positions of the GNSS receiver and GNSS satellites.
At middle latitudes this size has radius about 1500\textendash 2000
km. 
 AATR is sensitive
to the regional behavior of the ionosphere and can be used to identify
the conditions where a degradation in user performance is expected
\citep{sanz_2014}. 

The WTEC (Variations of Vertical TEC) index \citep{Voeykov_2016} allows
one to estimate the integral intensity of ionospheric disturbances
of various scales at single GNSS receiver.
The WTEC index represents the amplitude of variations
of vertical TEC filtered in a selected range of periods and averaged
over all the satellites visible by single GNSS receiver.
WTEC temporal resolution
is determined by the temporal resolution of the GNSS receiver and
usually about 30 s. 
Based on WTEC analysis \citep{Perevalova_2016} revealed the features
of large- and medium-scale TEC disturbances in mid-latitude and high-latitude
regions. Using WTEC for mid-scale disturbances, the ionospheric
effects of the Chelyabinsk meteorite fall on February 15, 2013 are
studied \citep{Voeykov_2016}.

Due to AATR and WTEC are calculated over approximately the same ionospheric
area but by different algorithms, their efficiency can be effectively
compared for their practical usage. In this paper, we performed an
analysis of AATR and WTEC indices to compare their efficiency for
studying mid-scale ionospheric disturbances in different conditions
and geographical locations. The analysis was carried out using the
data of high-latitude, equatorial and mid-latitude GNSS receivers
during 2014-2017.

\section{Calculation of AATR and WTEC }

The initial data for determining the AATR and WTEC indices are the
time series $I(t)$ of the Slant TEC along LOS and the time series
$\theta(t)$ of the GNSS satellite elevation angles. We calculated
STEC based on dual-frequency phase measurements at the GNSS receiver
by using standard method \citep{Hofmann_Wellenhof_2001}. STEC is measured
in TECU (Total Electron Content Unit): $1TECU=10^{16}el/m^{2}$. The
elevation angles are taken from the GNSS navigation messages. 

The values and dynamics of the AATR and WTEC indices can be significantly
distorted by gaps in phase measurements at GNSS receiver: missing
samples and cycle-slips. The both algorithms includes 
a preliminary detection and removal of the data with gaps 
before calculating the indices. The gaps can be associated with
the temporary loss of the satellite signal at GNSS receiver, as well
as with a small signal-to-noise ratio. The information about missing
samples is taken directly from the GNSS receiver data by using Lost
of Lock Indicator (LLI) \citep{Rinex_2015}. A cycle-slip is a change
of measured signal phase by an integer number of wavelengths. Several
techniques have been developed for detecting cycle-slips \citep{Hofmann_Wellenhof_2001,Juan_2018}.
In this paper, we detect the cycle-slip as the cases when the absolute
value of the second derivative of STEC for consecutive samples exceeded
a threshold value. The threshold is a constant determined individually
for each GNSS receiver used in the paper, it is based on the several
years of observations and is about $17-26\,TECU/min^{2}$. In the
paper we use only continuous data series without gaps, with a duration
of at least 20 minutes. To reduce the influence of the low-elevation
refraction effects we use only the data at high elevation angles exceeding
$15^{o}$ \citep{Voeykov_2016}. Selected data series have been used
for calculating AATR and WTEC. 

The algorithm for AATR calculation is described in detail in \citep{sanz_2014,Juan_2018}
and includes the following operations. Over the continuous data series
we calculate the rate of the STEC variation ($ROT_{k}^{j}$) between
two consecutive samples, received from j-th satellite during k-th
epoch ($t_{k}$). Instantaneous $AATR_{k}^{j}$ is calculated as $ROT_{k}^{j}$
multiplied by a slant factor $M_{k}^{j}$ to convert the slant TEC
to vertical TEC. In the approximation of a thin ionospheric layer
$M_{k}^{j}$ depends on the ionospheric layer height $h_{max}$ and
on the satellite elevation $\theta_{k}^{j}$. AATR index is calculated
as the Root Mean Square (RMS) of the instantaneous $AATR_{k}^{j}$
over all the visible satellites during the selected time interval
(1 hour). AATR index is measured in TECU/min with the temporal resolution
1 hour. 

The algorithm for WTEC calculation is described in detail in \citep{Voeykov_2016}
and includes the following operations. The continuous STEC series
$I_{k}^{j}(t)$ for each j-th satellite and k-th epoch are transformed
into vertical TEC using a slant factor $M_{k}^{j}$ in the thin spherical
ionosphere approximation. The obtained vertical TEC series are filtered
by a moving-average method to provide filtered series $dI_{k}^{j}(t)$
with periods less than T minutes. The average intensity of TEC oscillations
$A_{k}^{j}(t)$ is calculated as average absolute value of $dI_{k}^{j}(t)$
over T minutes period. $WTEC_{k}$ index for a given epoch $t_{k}$
is calculated as the weighted average of the $A_{k}^{j}(t)$ series
over all the visible satellites. Weighting function is specially chosen
to provide smooth estimate of WTEC values, not depending on number
of visible satellites. The resulting continuous $WTEC(t)$ series
reflect the average level of vertical TEC variations near the selected
station. WTEC is measured in TECU. Its temporal resolution is defined
by the temporal resolution of the GNSS receiver (30 seconds). In the
paper, we use T=10min that corresponds to mid-scale TEC disturbances.
It should be noted that the
period of TEC variations is closely related to the spatial scale of
the ionospheric disturbances: large-scale (LS) disturbances in the
ionosphere have periods of more than 30 minutes and size above 600
km, irregularities with periods of 5\textendash 20 minutes and sizes
of 50\textendash 600 km are usually considered as medium-scale (MS)
disturbances, small-scale (SC) disturbances have periods less than
5 min and sizes less than 30 km\citep{Hargreaves_1992}.

To make a joint analysis of AATR and WTEC indices, we produced the
software that calculates both the AATR and WTEC indices based on the
algorithms described in \citep{sanz_2014,Juan_2018,Voeykov_2016} and
use the same data sets without gaps for the calculations. The program
we developed is available at \citep{Berngardt_2019}. For the comparison
with AATR series the WTEC series are decimated to 1 hour temporal
resolution. As showed the analysis below, the released AATR calculation
algorithm produces the values similar to the results of \citep{sanz_2014}.
This allows us to verify our program and use it for the joint analysis
of WTEC and AATR indices.

\section{Experimental data }

\begin{figure}
\includegraphics[scale=0.6]{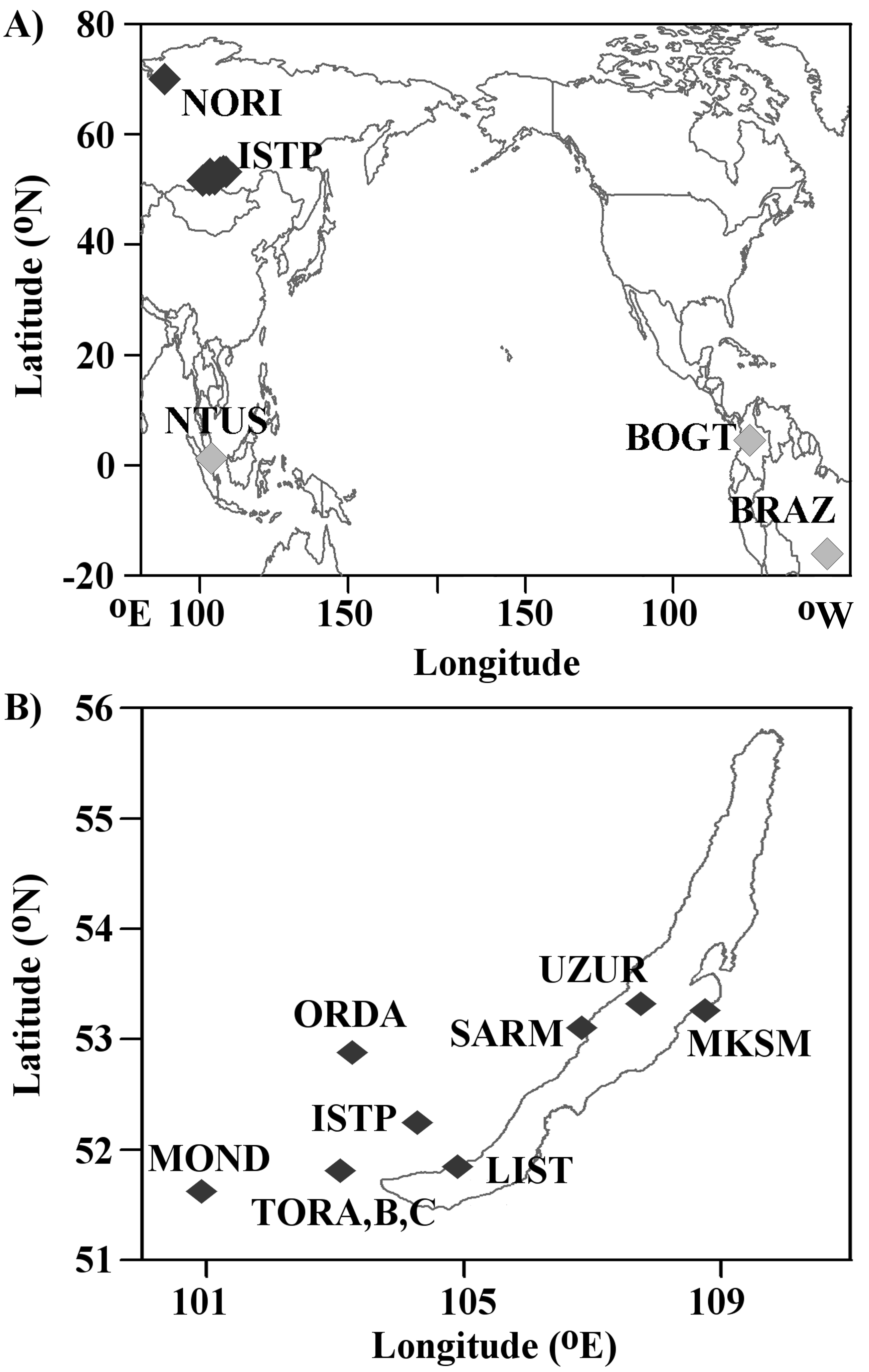}
\caption{Location of all GNSS stations used in this paper (A), ISTP SB RAS
stations near lake Baikal (B). The black and gray rhombs mark the GNSS
stations of ISTP SB RAS and IGS, correspondingly.}
\label{fig:fig1}
\end{figure}

For a comparative analysis of AATR and WTEC indices the data from
mid-latitude, high-latitude and low-latitude GNSS receivers are used.
Their locations are shown in Fig.\ref{fig:fig1}. The low-latitude
GNSS stations NTUS, BOGT and BRAZ are the part of International GNSS
Service (IGS) network \citep{IGS_data}. These low-latitude stations
are chosen to provide compatibility of this work with \citep{sanz_2014}.
Midlatitude ISTP SB RAS Baikal network (SibNet) of GNSS receivers
\citep{Ishin_2017} is shown in Fig.\ref{fig:fig1}B and includes 10
receivers (ORDA, UZUR, SARM, MKSM, ISTP, LIST, MOND, TORA, TORB, TORC).
In Fig.\ref{fig:fig1}A SibNet is marked as 'ISTP'. Also for analysis
we use the data of high-latitude NORI receiver of ISTP SB RAS. Below
we present results of comparative statistical analysis of AATR and
WTEC dynamics over the period 2014-2017 using GNSS data and the algorithms
described above. We also present results of regression analysis of
AATR and WTEC with parameters of solar, geomagnetic and seismic activity
as well as with background total electron content. We also demonstrate
AATR and WTEC sensitivity to strong and weak ionospheric disturbances
generated by geomagnetic storms and earthquakes correspondingly. 

The following solar and geophysical data were used for the analysis: 
\begin{itemize}
\item the geomagnetic indices Kp, Dst, SYM-H obtained from Kyoto World Data
Center for Geomagnetism \citep{Kyoto_WDC_data} and OMNIWeb database\citep{OMNIWeb_data}; 
\item the vertical total electron content (VTEC), characterizing the background
ionospheric conditions obtained from Global Ionosphere Maps (GIM)
produced by Center for Orbit Determination in Europe (CODE). The CODE
GIMs were obtained from the Crustal Dynamics Data Information System
(CDDIS) at Goddard Space Flight Center, NASA\citep{CDDIS_data}; 
\item the F10.7 index of solar radiation at 10.7 cm wavelength, maximal
over 1 hour and obtained from OMNIWeb database \citep{OMNIWeb_data}; 
\item the vertical seismic variation (dZ) according to the data of Talaya
seismic station (TLY) maximal over 1 hour and obtained from the IRIS
(Incorporated Research Institutions for Seismology) database\citep{IRIS_data}.
TLY seismic station is located near the GNSS receiver TORA. 
\end{itemize}

\begin{figure}
\includegraphics[scale=0.25]{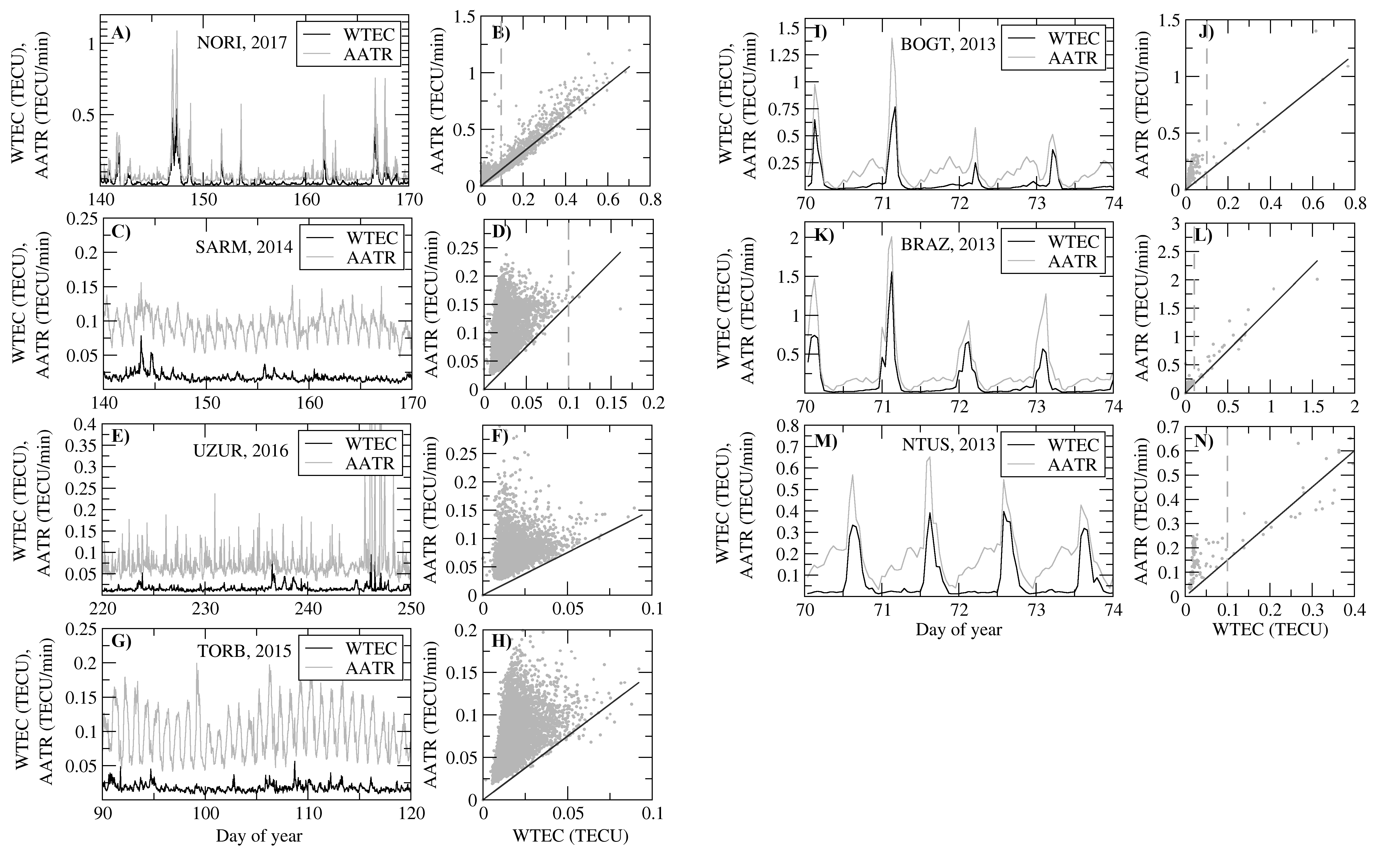}
\caption{Comparison of AATR (gray line) and WTEC (black line) indices dynamics
based on the data from various stations - high-latitude (A-B), mid-latitude
(C-H) and low-latitude (I-N) ones. Left panels correspond to the dynamics
of indices and right panels correspond to regression dependence of
AATR on WTEC. Gray points are experimental points, the solid line
corresponds to the regression dependence (1). Vertical dashed line
limits the region WTEC<0.1TECU.}
\label{fig:fig2}
\end{figure}

Statistical analysis of the indices Fig.\ref{fig:fig2} shows WTEC
and AATR indices at high-latitude (Fig.\ref{fig:fig2}A-B), mid-latitude
(Fig.\ref{fig:fig2}C-H) and equatorial (Fig.\ref{fig:fig2}I-N) GNSS
stations. For the analysis of equatorial station data we use the days
070-074 of 2013, studied in detail in \citep{sanz_2014} over the identical
set of low-latitude stations. Comparison of Fig.\ref{fig:fig2}I-N
with results of \cite[,fig.10]{sanz_2014} shows a good agreement
between them and allows us to verify our program and use its results
for the joint analysis of the WTEC and AATR indices. One can see in
Fig.\ref{fig:fig2} that for relatively large WTEC>0.1TECU, the AATR
and WTEC indices correlate well with each other (Fig.\ref{fig:fig2}B,J,L,N)
and their relation can be described by the proportion: 

\begin{equation}
AATR(t)[TECU/min]=1.5[min^{-1}]\cdot WTEC(t)[TECU]\label{eq:1}
\end{equation}

For small WTEC<0.1TECU, the correlation between AATR and WTEC is less
obvious (Fig.\ref{fig:fig2}D,F,H), and the relationship between the
indices can be represented as: 

\begin{equation}
AATR(t)[TECU/min]\geq1.5[min^{-1}]\cdot WTEC(t)[TECU]\label{eq:2}
\end{equation}

This is an indirect sign that AATR index can be considered as a superposition
of the term proportional to WTEC index and additional small term of
the order of 0.1TECU, that is significant only for mid-latitude observations
where WTEC values are regularly small. The region WTEC<0.1TECU is
shown in Fig.\ref{fig:fig2}B,J,L,N by dashed line. As one can see
in Fig.\ref{fig:fig2}, large WTEC values are more often observed
at high-latitude and equatorial stations. These latitudes are characterized
by higher level of ionospheric disturbances. Therefore, at these stations
during strong ionospheric disturbances (for example, generated by
geomagnetic storms) one can use both AATR and WTEC indices with equal
efficiency. At mid-latitude stations, where the disturbance level
is usually lower, one should use a more sensitive index of these two.

\section{Sensitivity of the indices to space and surface disturbances}

Fig.\ref{fig:fig2} shows that during strong disturbances (corresponding
to high AATR and WTEC values) at high-latitudes and low-latitudes
AATR and WTEC indices correlate well and are nearly proportional to
each other, and using them for analysis is equivalent. For a more
detailed analysis of these indices at mid- and high-latitudes, we
use the large data set of ISTP SB RAS GNSS stations: mid-latitude
SibNet network near the lake Baikal (10 stations) and high-latitude
NORI station in the Arctic region. The Baikal region is characterized
by significant seismic activity, which makes it possible to estimate
the contribution to AATR and WTEC indices the various sources of disturbances,
both from the space and from the ground. In Fig.\ref{fig:fig2}C,E,G,I,M
one can see that the AATR index at mid- and low latitudes has a high
level of regular variations. A preliminary spectral analysis shows
the presence of a substantial diurnal component in the AATR index,
which is less significant in the WTEC index. To study the sources
of regular and irregular variations of the indices we estimate regression
dependence of WTEC and AATR indices on various parameters of the solar,
geomagnetic and seismic activity as well as on background ionospheric
conditions as the following: 

\begin{equation}
\begin{array}{l}
WTEC_{fit}(t)=A_{0}+\sum_{i=1}^{4}F_{i}(t)=A_{0}+\sum_{i=1}^{4}A_{i}I_{i}(t)\\
AATR_{fit}(t)=B_{0}+\sum_{i=1}^{4}G_{i}(t)=B_{0}+\sum_{i=1}^{4}B_{i}I_{i}(t)
\end{array}\label{eq:3}
\end{equation}

Unknown regression coefficients $A_{i},B_{i}$ are calculated by the
least squares method to provide the best fit of the experimental measurements
($WTEC(t),AATR(t)$) by the models (\ref{eq:3}): 

\begin{equation}
\begin{array}{l}
\int_{-\infty}^{\infty}\left(WTEC(t)-WTEC_{fit}(t)\right)^{2}dt=min\\
\int_{-\infty}^{\infty}\left(AATR(t)-AATR_{fit}(t)\right)^{2}dt=min
\end{array}\label{eq:4}
\end{equation}

The following time series with a temporal resolution 1 hour were used
for regression analysis of WTEC and AATR (\ref{eq:3}-\ref{eq:4}):
$I_{1}(t)$ is the geomagnetic index Kp; $I_{2}(t)$ is VTEC, characterizing
the background ionospheric conditions obtained from Global Ionosphere
Maps (GIM); $I_{3}(t)$ is F10.7 index of solar activity; $I_{4}(t)$
is the vertical seismic variation (dZ) according to the data of Talaya
seismic station (TLY). When processing to bring all the parameters
to the same temporal resolution, rare data (Kp index) are approximated
by a step function between measurements, for frequent data the maximum
value over 1 hour is taken.

We calculated the relative contribution $\alpha_{i},\beta_{i}$ of
each time series to AATR and WTEC indices as: 

\begin{equation}
\begin{array}{l}
\alpha_{i}=\frac{\int_{-\infty}^{\infty}\left|F_{i}(t)\right|dt}{\sum_{j=1}^{4}\int_{-\infty}^{\infty}\left|F_{j}(t)\right|dt}\\
\beta_{i}=\frac{\int_{-\infty}^{\infty}\left|G_{i}(t)\right|dt}{\sum_{j=1}^{4}\int_{-\infty}^{\infty}\left|G_{j}(t)\right|dt}
\end{array}\label{eq:5}
\end{equation}

The parameters $\alpha_{i},\beta_{i}$ are non-negative and dimensionless,
and the bigger $\alpha_{i},\beta_{i}$ the greater the contribution
to the index value is made by the disturbance component $I_{i}(t)$.
Thus, the analysis of the parameters $\alpha_{i},\beta_{i}$ obtained
as a result of fitting eq.(\ref{eq:4}), allows us to identify the
main components that regularly affect the behavior of AATR and WTEC
indices and to make a comparative analysis of the contribution of
each source of disturbances to WTEC and AATR. The calculated parameters
$\alpha_{i},\beta_{i}$ are shown in the Table 1.

\begin{table}[!htbp]
\caption{Sensitivity of AATR and WTEC indices to geomagnetic, solar and seismic
disturbances according to ISTP SB RAS GNSS stations}
\centering
\begin{tabular}{|c|c|c|c|c|}
\toprule
Mode &  \multicolumn{2}{|c|}{Average over 10 mid-latitude SibNet stations} & \multicolumn{2}{|c|}{High-latitude station NORI}\\
\midrule
{}   & $\alpha_{i}$(WTEC) & $\beta_{i}$(AATR) & $\alpha_{i}$(WTEC) & $\beta_{i}$(AATR)\\
\hline 
Kp & 0.011 & 0.002 & 0.734 & 0.957\\
\hline 
VTEC & 0.874 & 0.995 & 0.262 & 0.035\\
\hline 
F10.7 & 0.093 & 0.002 & 0.004 & 0.008\\
\hline 
dZ & 0.022 & 0.002 & 0.000 & 0.000\\
\bottomrule
\end{tabular}
\label{tab:1}
\end{table}

The Table \ref{tab:1} shows that at mid-latitude stations the influence
of the regular VTEC on WTEC is less than its influence on AATR. WTEC
dependence on intensity of seismic and geomagnetic disturbances is
more pronounced than AATR dependency. It should be noted that the
small influence of the Kp level at mid-latitude stations may be caused
by the weakness of the effects of geomagnetic disturbances at mid-latitudes.
The small influence of the seismic activity may be caused by relatively
weak seismic activity of the region during the period of observations
(maximal magnitude of the seismic events in near-Baikal region was
M5.4). At high-latitude NORI station, the dependence on geomagnetic
disturbance prevails others both for AATR and for WTEC, which confirms
well with the conclusions of \citep{Juan_2018} about the possibility
of using the AATR index for analyzing ionospheric disturbances at
high latitudes during geomagnetic disturbances. The smallness of effects
of VTEC at high latitudes can be related with smaller diurnal variation
of VTEC. 

\section{Case comparison of AATR and WTEC during strong and weak disturbances }

\begin{figure}
\includegraphics[scale=0.25]{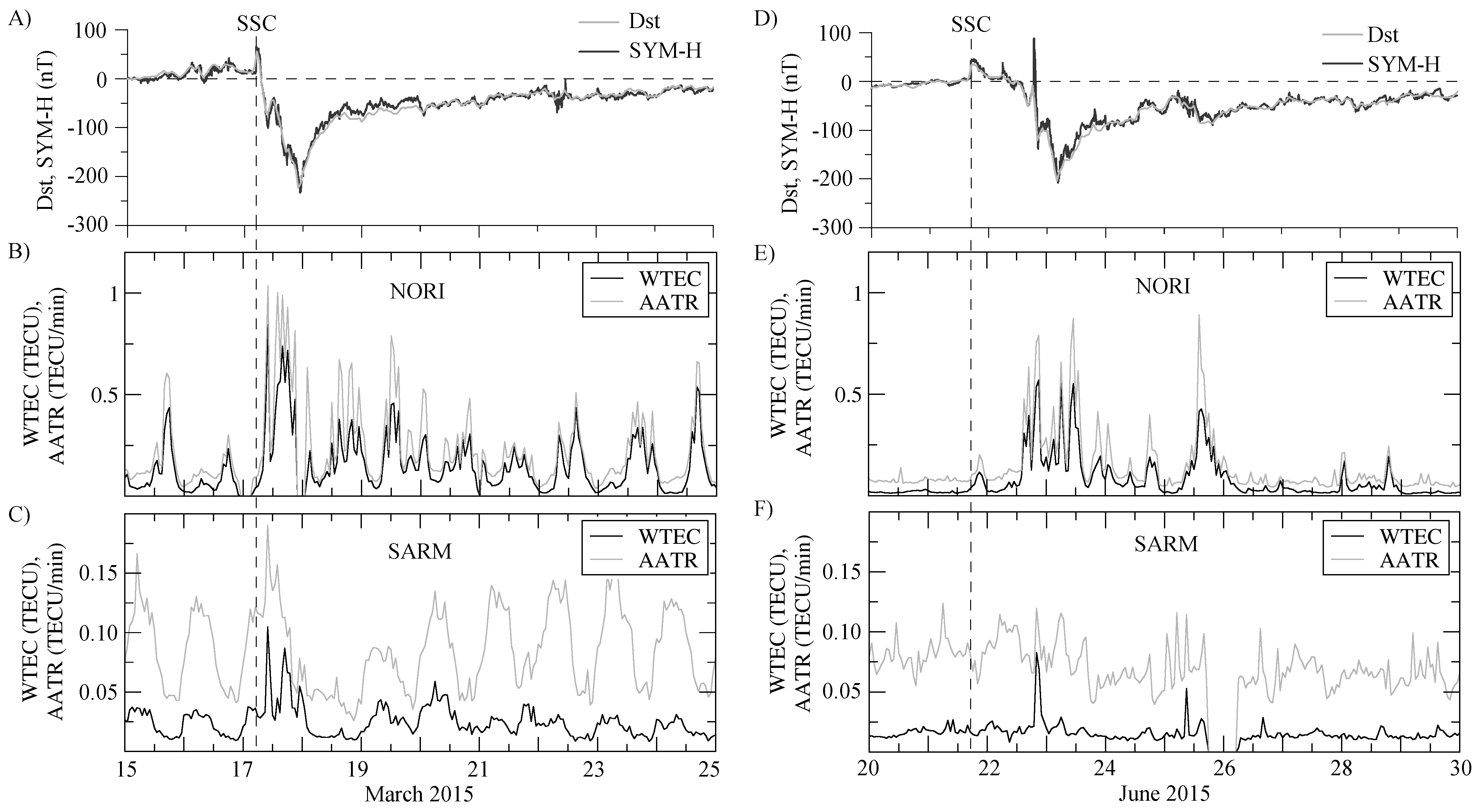}
\caption{AATR and WTEC dynamics at high-latitude (NORI) and mid-latitude (SARM)
ISTP SB RAS GNSS-stations during March 15-25, 2015 (A-C) and June
20-30, 2015 (D-F) geomagnetic storms. The vertical dashed line marks
the SSC. A, D) - Dst and SYM-H; B,E) - WTEC and AATR at NORI; C,F)
- WTEC and AATR at SARM.}
\label{fig:fig3}
\end{figure}

To illustrate the sensitivity of AATR and WTEC indices to different
disturbances, their responses to some geomagnetic storms and earthquakes
were calculated. Geomagnetic storms cause strong disturbances in the
ionosphere. Detection and study of these disturbances are one of the
practically important applications of indices. Geomagnetic storms
on March 17-25 and June 21-30, 2015 are among the strongest storms
during 24 solar activity cycle\citep{Liu_2015}. To characterize the
geomagnetic field disturbance, we used the Dst (resolution of 1 hour)
and SYM-H (resolution of 1 minute) geomagnetic indices. Fig.\ref{fig:fig3}A
shows the SYM-H and Dst dynamics during the March 17-25, 2015 geomagnetic
storm. According to SYM-H, the storm sudden commencement (SSC) was
registered at 04:45 UT on March 17. The maximum of the storm was observed
at 22:47UT on March 17, when the geomagnetic indices reached their
minimum values: SYM-H=-234 nT, Dst=-223 nT (Fig.\ref{fig:fig3}A).
The geomagnetic field disturbance during the June 21-30, 2015 storm
is shown in Fig.\ref{fig:fig3}D. SYM-H and Dst dynamics shows SSC
at 16:45UT on June 21, 2015 and several peaks during the storm main
phase. The main minimum of SYM-H (-206 nT) and Dst (-204 nT) was registered
on June 23 at 04:38 UT, after which the long (lasted until June 30)
storm recovery phase began (Fig.\ref{fig:fig3}D). Fig.\ref{fig:fig3}B,E
and Fig.\ref{fig:fig3}C,F demonstrate the behavior of the AATR and
WTEC indices at high-latitude NORI station and at mid-latitude SARM
station. As we can see in (Fig.\ref{fig:fig3}B,E), WTEC and AATR
correlate well during storm time at high-latitudes. At mid-latitudes
(Fig.\ref{fig:fig3}C,F), the storm effect is more pronounced in WTEC
data; in the AATR data the storm generated disturbance is small compared
to the quasi-regular daily AATR dynamics.

\begin{figure}
\includegraphics[scale=0.35]{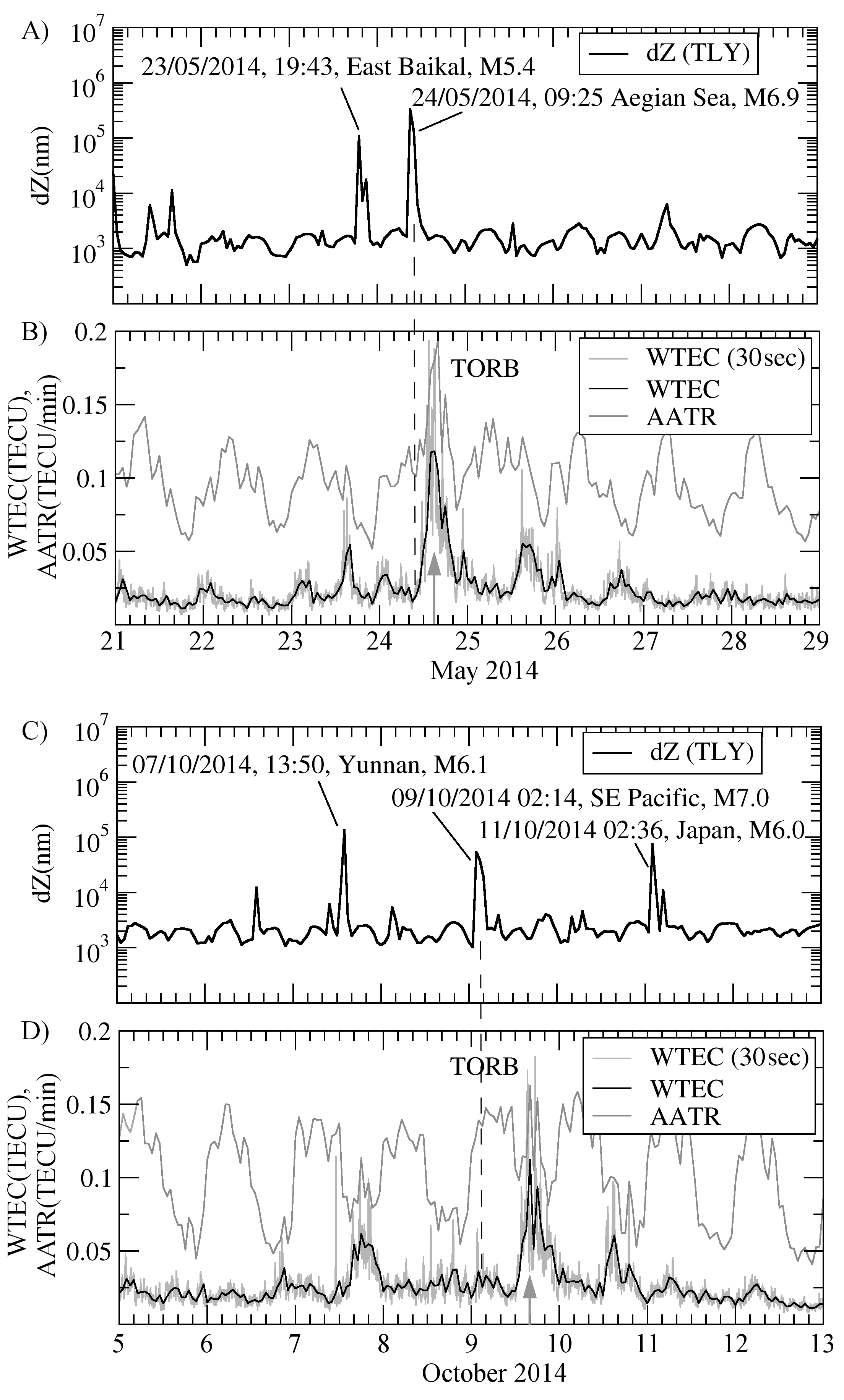}
\caption{AATR and WTEC dynamics during seismic disturbances. A,C) - data of
the Talaya seismic station (TLY). B,D) - TORB GNSS-receiver data.
The light gray line corresponds to WTEC with 30-second temporal resolution,
the black and dark gray lines mark, respectively, WTEC and AATR with 1 hour
temporal resolution. Vertical dashed line marks the earthquake vibrations
received at TLY station. Arrow marks the ionospheric response possibly
related with it.}
\label{fig:fig4}
\end{figure}

Earthquakes usually produce less powerful ionospheric disturbances
of various types. The basic sources of ionospheric disturbances are
internal atmospheric waves generated by seismic vibrations near epicenter
\citep{Astafyeva_2013} and shock acoustic waves generated by surface
seismic waves near observation point \citep{Berngardt_2017}. Fig.\ref{fig:fig4}B,D
demonstrate AATR and WTEC behavior during the seismic vibrations observed
at TLY station and shown in Fig.\ref{fig:fig4}A,C. Both WTEC and
AATR allow one to detect powerful disturbances, for example after
the earthquake in the Aegean Sea on 24/05/2014 09:25UT, about 5800
km from SibNet (Fig.\ref{fig:fig4}A-B). On the other hand, WTEC make
it possible to register weaker effects than AATR, for example, disturbances
observed several hours after the Pacific earthquake of 09/10/2014
02:14UT, about 16500 km from SibNet (Fig.\ref{fig:fig4}C-D). Ionospheric
response in both cases (marked at Fig.\ref{fig:fig4}B,D by arrows)
can be related with the passage of internal atmospheric waves, which
can be justified by the presence of significant several hours delay
between seismic observation (dashed line) and ionospheric observation
(arrow). This delay between them corresponds to atmospheric wave propagation
velocity (about 250-350m/s). In the AATR data these disturbances are
also present, but less pronounced due to the presence of the diurnal
variation of the index. The effect of the earthquake on 23/05/2014
19:43UT at East Baikal looks not observed, which is associated, apparently,
with earthquake insufficient magnitude \citep{PEREVALOVA_2014}. Thus,
the comparison of AATR and WTEC dynamics confirms our conclusion about
the higher sensitivity of the WTEC index to weak disturbances as compared
to AATR and the lesser influence of the background electron dynamics
to WTEC.

\section{Conclusion}

An important task of monitoring ionospheric disturbances at mid latitudes
using single GNSS-receivers is to construct sensitive and effective
indices describing the intensity of ionospheric disturbances. One
of the widely used indices today is the ROTI index \citep{Pi_1997}
and AATR index - its average satellite-receiver beam value averaged
over 1 hour \citep{sanz_2014,Juan_2018}. The indices are often used
to analyze the high-latitude and equatorial ionosphere, but usually
are not used at mid-latitudes. The aim of this paper was to study
the efficiency of using the WTEC index obtained with a 30-second time
resolution \citep{Voeykov_2016} for analyzing disturbances in the
mid-latitude ionosphere based on long-term observations at the complex
of mid-latitude GNSS receivers of the ISTP SB RAS and its comparative
analysis with the AATR index. 

A case comparative analysis of AATR and WTEC indices was performed
based on selected GNSS receivers. It is shown that at high and low
latitudes at high levels of ionospheric disturbance (when WTEC > 0.1TECU),
AATR is proportional to WTEC with a factor of $1.5min^{-1}$. At mid-latitudes
at small levels of ionospheric disturbance (when WTEC <0.1TECU), this
proportionality is violated. In this case AATR level becomes higher
than expected. 

A comparative statistical analysis of AATR and WTEC indices was performed
based on ISTP SB RAS mid- and high-latitude GNSS receivers during
period 2014-2017. A regression analysis was made with parameters of
solar, geomagnetic and seismic activity as well as with background
total electron content. It is shown that the contribution of daily
dynamics of the background ionosphere to the AATR index is higher
than to the WTEC index. This leads to a higher sensitivity of the
WTEC index to different disturbances. This also leads to violating
the proportionality between WTEC and AATR indices at low levels of
ionospheric disturbances. It is shown that at high latitudes the dynamics
of the WTEC and AATR indices correlate significantly with the level
of geomagnetic disturbance Kp. At mid-latitudes, the contribution
of solar radiation variations (F10.7 index) and seismic activity exceeds
the influence of Kp variations. 

The analysis of strong and weak events associated with geomagnetic
disturbances and seismic disturbances demonstrated the higher sensitivity
of the WTEC index than AATR at mid latitudes to ionospheric disturbances.
Thus, at WTEC <0.1TECU, the use of the WTEC index looks more accurate;
at high disturbance levels, the WTEC and AATR indices are almost proportional
to each other so their use at high and low latitudes looks equivalent.
The program for calculating WTEC indices, used in the paper is put
into open access \citep{Berngardt_2019}.

\section*{Acknowledgments}

The authors are grateful to Center for Orbit Determination in Europe
for GIM data available on the website ftp://cddis.gsfc.nasa.gov:21/gnss/products/ionex.
We are grateful to IRIS/IDA Seismic Network (II), Global Seismograph
Network (GSN - IRIS/USGS) (GSN,IU), SY - Synthetic Seismograms Network
(SY) for providing seismic data. 
SibNet from Angara Center for Common Use of scientific equipment (http://ckp-rf.ru/ckp/3056/) is operated under budgetary funding of Basic Research program II.12.
The work was done under support of
the joint grant \#17-45-388072r\_a of RFBR and Goverment of Irkutsk Region, Russia.


\begin{thebibliography}{44}
\expandafter\ifx\csname natexlab\endcsname\relax\def\natexlab#1{#1}\fi
\providecommand{\url}[1]{\texttt{#1}}
\providecommand{\href}[2]{#2}
\providecommand{\path}[1]{#1}
\providecommand{\DOIprefix}{doi:}
\providecommand{\ArXivprefix}{arXiv:}
\providecommand{\URLprefix}{URL: }
\providecommand{\Pubmedprefix}{pmid:}
\providecommand{\doi}[1]{\href{http://dx.doi.org/#1}{\path{#1}}}
\providecommand{\Pubmed}[1]{\href{pmid:#1}{\path{#1}}}
\providecommand{\bibinfo}[2]{#2}
\ifx\xfnm\relax \def\xfnm[#1]{\unskip,\space#1}\fi
\bibitem[{Afraimovich et~al.(2009)Afraimovich, Edemskiy, Voeykov, Yasyukevich
  and Zhivetiev}]{Afraimovich_2009}
\bibinfo{author}{Afraimovich, E.}, \bibinfo{author}{Edemskiy, I.},
  \bibinfo{author}{Voeykov, S.}, \bibinfo{author}{Yasyukevich, Y.},
  \bibinfo{author}{Zhivetiev, I.}, \bibinfo{year}{2009}.
\newblock \bibinfo{title}{Spatio-temporal structure of the wave packets
  generated by the solar terminator}.
\newblock \bibinfo{journal}{Advances in Space Research} \bibinfo{volume}{44},
  \bibinfo{pages}{824 -- 835}.
\newblock \DOIprefix\doi{https://doi.org/10.1016/j.asr.2009.05.017}.
\bibitem[{{Afraimovich, E.L.} et~al.(2013){Afraimovich, E.L.}, {Astafyeva,
  E.I.}, {Demyanov, V.V.}, {Edemskiy, I.K.}, {Gavrilyuk, N.S.}, {Ishin, A.B.},
  {Kosogorov, E.A.}, {Leonovich, L.A.}, {Lesyuta, O.S.}, {Palamartchouk, K.S.},
  {Perevalova, N.P.}, {Polyakova, A.S.}, {Smolkov, G.Y.}, {Voeykov, S.V.},
  {Yasyukevich, Y.V.} and {Zhivetiev, I.V.}}]{Afraimovich_2013}
\bibinfo{author}{{Afraimovich, E.L.}}, \bibinfo{author}{{Astafyeva, E.I.}},
  \bibinfo{author}{{Demyanov, V.V.}}, \bibinfo{author}{{Edemskiy, I.K.}},
  \bibinfo{author}{{Gavrilyuk, N.S.}}, \bibinfo{author}{{Ishin, A.B.}},
  \bibinfo{author}{{Kosogorov, E.A.}}, \bibinfo{author}{{Leonovich, L.A.}},
  \bibinfo{author}{{Lesyuta, O.S.}}, \bibinfo{author}{{Palamartchouk, K.S.}},
  \bibinfo{author}{{Perevalova, N.P.}}, \bibinfo{author}{{Polyakova, A.S.}},
  \bibinfo{author}{{Smolkov, G.Y.}}, \bibinfo{author}{{Voeykov, S.V.}},
  \bibinfo{author}{{Yasyukevich, Y.V.}}, \bibinfo{author}{{Zhivetiev, I.V.}},
  \bibinfo{year}{2013}.
\newblock \bibinfo{title}{A review of gps/glonass studies of the ionospheric
  response to natural and anthropogenic processes and phenomena}.
\newblock \bibinfo{journal}{J. Space Weather Space Clim.} \bibinfo{volume}{3},
  \bibinfo{pages}{A27}.
\newblock \DOIprefix\doi{10.1051/swsc/2013049}.
\bibitem[{Astafyeva et~al.(2013)Astafyeva, Shalimov, Olshanskaya and
  Lognonn\'{e}}]{Astafyeva_2013}
\bibinfo{author}{Astafyeva, E.}, \bibinfo{author}{Shalimov, S.},
  \bibinfo{author}{Olshanskaya, E.}, \bibinfo{author}{Lognonn\'{e}, P.},
  \bibinfo{year}{2013}.
\newblock \bibinfo{title}{{Ionospheric response to earthquakes of different
  magnitudes: Larger quakes perturb the ionosphere stronger and longer}}.
\newblock \bibinfo{journal}{Geophysical Research Letters} \bibinfo{volume}{40},
  \bibinfo{pages}{1675--1681}.
\newblock \URLprefix
  \url{https://agupubs.onlinelibrary.wiley.com/doi/abs/10.1002/grl.50398},
  \DOIprefix\doi{10.1002/grl.50398}.
\bibitem[{Astafyeva and Shults(2018)}]{Astafyeva_2018}
\bibinfo{author}{Astafyeva, E.}, \bibinfo{author}{Shults, K.},
  \bibinfo{year}{2018}.
\newblock \bibinfo{title}{Ionospheric gnss imagery of seismic source:
  Possibilities, difficulties, and challenges}.
\newblock \bibinfo{journal}{Journal of Geophysical Research: Space Physics}
  \bibinfo{volume}{0}.
\newblock \DOIprefix\doi{10.1029/2018ja026107}.
\bibitem[{B\'{e}niguel et~al.(2004)B\'{e}niguel, Forte, Radicella, Strangeways,
  Gherm and Zernov}]{Beniguel_2004}
\bibinfo{author}{B\'{e}niguel, Y.}, \bibinfo{author}{Forte, B.},
  \bibinfo{author}{Radicella, S.M.}, \bibinfo{author}{Strangeways, H.J.},
  \bibinfo{author}{Gherm, V.E.}, \bibinfo{author}{Zernov, N.N.},
  \bibinfo{year}{2004}.
\newblock \bibinfo{title}{{Scintillations effects on satellite to Earth links
  for telecommunication and navigation purposes}}.
\newblock \bibinfo{journal}{Annals of Geophysics} \bibinfo{volume}{47}.
\newblock \DOIprefix\doi{10.4401/ag-3293}.
\bibitem[{Berngardt et~al.(2015)Berngardt, Perevalova, Dobrynina, Kutelev,
  Shestakov, Bakhtiarov, Kusonsky, Zagretdinov and Zherebtsov}]{Berngardt_2015}
\bibinfo{author}{Berngardt, O.I.}, \bibinfo{author}{Perevalova, N.P.},
  \bibinfo{author}{Dobrynina, A.A.}, \bibinfo{author}{Kutelev, K.A.},
  \bibinfo{author}{Shestakov, N.V.}, \bibinfo{author}{Bakhtiarov, V.F.},
  \bibinfo{author}{Kusonsky, O.A.}, \bibinfo{author}{Zagretdinov, R.V.},
  \bibinfo{author}{Zherebtsov, G.A.}, \bibinfo{year}{2015}.
\newblock \bibinfo{title}{Toward the azimuthal characteristics of ionospheric
  and seismic effects of "chelyabinsk" meteorite fall according to the data
  from coherent radar, gps, and seismic networks}.
\newblock \bibinfo{journal}{Journal of Geophysical Research: Space Physics}
  \bibinfo{volume}{120}, \bibinfo{pages}{10,754--10,771}.
\newblock \DOIprefix\doi{10.1002/2015JA021549}.
\bibitem[{Berngardt et~al.(2017)Berngardt, Perevalova, Podlesnyi, Kurkin and
  Zherebtsov}]{Berngardt_2017}
\bibinfo{author}{Berngardt, O.I.}, \bibinfo{author}{Perevalova, N.P.},
  \bibinfo{author}{Podlesnyi, A.V.}, \bibinfo{author}{Kurkin, V.I.},
  \bibinfo{author}{Zherebtsov, G.A.}, \bibinfo{year}{2017}.
\newblock \bibinfo{title}{{Vertical midscale ionospheric disturbances caused by
  surface seismic waves based on Irkutsk chirp ionosonde data in 2011-2016}}.
\newblock \bibinfo{journal}{Journal of Geophysical Research: Space Physics}
  \bibinfo{volume}{122}, \bibinfo{pages}{4736--4754}.
\newblock \URLprefix
  \url{https://agupubs.onlinelibrary.wiley.com/doi/abs/10.1002/2016JA023511},
  \DOIprefix\doi{10.1002/2016JA023511}.
\bibitem[{CDDIS(2019)}]{CDDIS_data}
\bibinfo{author}{CDDIS}, \bibinfo{year}{2019}.
\newblock \bibinfo{title}{Cddis database}.
\newblock \URLprefix \url{ftp://cddis.gsfc.nasa.gov:21/gnss/products/ionex}.
\bibitem[{Cherniak et~al.(2018)Cherniak, Krankowski and
  Zakharenkova}]{Cherniak2018}
\bibinfo{author}{Cherniak, I.}, \bibinfo{author}{Krankowski, A.},
  \bibinfo{author}{Zakharenkova, I.}, \bibinfo{year}{2018}.
\newblock \bibinfo{title}{Roti maps: a new igs ionospheric product
  characterizing the ionospheric irregularities occurrence}.
\newblock \bibinfo{journal}{GPS Solutions} \bibinfo{volume}{22},
  \bibinfo{pages}{69}.
\newblock \DOIprefix\doi{10.1007/s10291-018-0730-1}.
\bibitem[{Gulyaeva et~al.(2013)Gulyaeva, Arikan, Hernandez-Pajares and
  Stanislawska}]{GULYAEVA_2013}
\bibinfo{author}{Gulyaeva, T.}, \bibinfo{author}{Arikan, F.},
  \bibinfo{author}{Hernandez-Pajares, M.}, \bibinfo{author}{Stanislawska, I.},
  \bibinfo{year}{2013}.
\newblock \bibinfo{title}{Gim-tec adaptive ionospheric weather assessment and
  forecast system}.
\newblock \bibinfo{journal}{Journal of Atmospheric and Solar-Terrestrial
  Physics} \bibinfo{volume}{102}, \bibinfo{pages}{329 -- 340}.
\newblock \DOIprefix\doi{https://doi.org/10.1016/j.jastp.2013.06.011}.
\bibitem[{Gulyaeva and Stanislawska(2008)}]{Gulyaeva_and_Stanislavswska_2008}
\bibinfo{author}{Gulyaeva, T.L.}, \bibinfo{author}{Stanislawska, I.},
  \bibinfo{year}{2008}.
\newblock \bibinfo{title}{Derivation of a planetary ionospheric storm index}.
\newblock \bibinfo{journal}{Annales Geophysicae} \bibinfo{volume}{26},
  \bibinfo{pages}{2645--2648}.
\newblock \DOIprefix\doi{10.5194/angeo-26-2645-2008}.
\bibitem[{Hargreaves(1992)}]{Hargreaves_1992}
\bibinfo{author}{Hargreaves, J.}, \bibinfo{year}{1992}.
\newblock \bibinfo{title}{The solar-terrestrial environment}.
\newblock \bibinfo{publisher}{Cambridge University Press}.
\newblock \DOIprefix\doi{10.1017/cbo9780511628924}.
\bibitem[{Ho et~al.(1998)Ho, Mannucci, Sparks, Pi, Lindqwister, Wilson, Iijima
  and Reyes}]{Ho_1998}
\bibinfo{author}{Ho, C.M.}, \bibinfo{author}{Mannucci, A.J.},
  \bibinfo{author}{Sparks, L.}, \bibinfo{author}{Pi, X.},
  \bibinfo{author}{Lindqwister, U.J.}, \bibinfo{author}{Wilson, B.D.},
  \bibinfo{author}{Iijima, B.A.}, \bibinfo{author}{Reyes, M.J.},
  \bibinfo{year}{1998}.
\newblock \bibinfo{title}{Ionospheric total electron content perturbations
  monitored by the gps global network during two northern hemisphere winter
  storms}.
\newblock \bibinfo{journal}{Journal of Geophysical Research: Space Physics}
  \bibinfo{volume}{103}, \bibinfo{pages}{26409--26420}.
\newblock \DOIprefix\doi{10.1029/98JA01237}.
\bibitem[{Hofmann-Wellenhof et~al.(2001)Hofmann-Wellenhof, Lichtenegger and
  Collins}]{Hofmann_Wellenhof_2001}
\bibinfo{author}{Hofmann-Wellenhof, B.}, \bibinfo{author}{Lichtenegger, H.},
  \bibinfo{author}{Collins, J.}, \bibinfo{year}{2001}.
\newblock \bibinfo{title}{Global Positioning System}.
\newblock \bibinfo{publisher}{Springer Vienna}.
\newblock \DOIprefix\doi{10.1007/978-3-7091-6199-9}.
\bibitem[{IGS(2019)}]{IGS_data}
\bibinfo{author}{IGS}, \bibinfo{year}{2019}.
\newblock \bibinfo{title}{Igs database}.
\newblock \URLprefix \url{ftp://garner.ucsd.edu:21/pub/rinex/}.
\bibitem[{IGS and RTCM-SC104(2015)}]{Rinex_2015}
\bibinfo{author}{IGS}, \bibinfo{author}{RTCM-SC104}, \bibinfo{year}{2015}.
\newblock \bibinfo{title}{Rinex. the receiver independent exchange format.
  version 3.03}.
\newblock \URLprefix \url{ftp://igs.org/pub/data/format/rinex303.pdf}.
\bibitem[{IRIS(2019)}]{IRIS_data}
\bibinfo{author}{IRIS}, \bibinfo{year}{2019}.
\newblock \bibinfo{title}{Iris database}.
\newblock \URLprefix \url{http://www.ds.iris.edu/ds/}.
\bibitem[{Ishin et~al.(2017)Ishin, Perevalova, Voeykov and
  Khakhinov}]{Ishin_2017}
\bibinfo{author}{Ishin, A.}, \bibinfo{author}{Perevalova, N.},
  \bibinfo{author}{Voeykov, S.}, \bibinfo{author}{Khakhinov, V.},
  \bibinfo{year}{2017}.
\newblock \bibinfo{title}{{First results of registering ionospheric
  disturbances obtained with SibNet network of GNSS receivers in active space
  experiments}}.
\newblock \bibinfo{journal}{Solar-Terrestrial Physics} \bibinfo{volume}{3},
  \bibinfo{pages}{74--82}.
\newblock \DOIprefix\doi{10.12737/stp-34201708}.
\bibitem[{Jacobsen and Andalsvik(2016)}]{Jacobsen_2016}
\bibinfo{author}{Jacobsen, K.}, \bibinfo{author}{Andalsvik, Y.},
  \bibinfo{year}{2016}.
\newblock \bibinfo{title}{Overview of the 2015 st. patrick\'{}s day storm and
  its consequences for rtk and ppp positioning in norway}.
\newblock \bibinfo{journal}{J. Space Weather Space Clim.} \bibinfo{volume}{6},
  \bibinfo{pages}{A9}.
\newblock \URLprefix \url{https://doi.org/10.1051/swsc/2016004},
  \DOIprefix\doi{10.1051/swsc/2016004}.
\bibitem[{Jacobsen and D\"ahnn(2014)}]{Jacobsen_2014}
\bibinfo{author}{Jacobsen, K.}, \bibinfo{author}{D\"ahnn, M.},
  \bibinfo{year}{2014}.
\newblock \bibinfo{title}{Statistics of ionospheric disturbances and their
  correlation with gnss positioning errors at high latitudes}.
\newblock \bibinfo{journal}{J. Space Weather Space Clim.} \bibinfo{volume}{4},
  \bibinfo{pages}{A27}.
\newblock \DOIprefix\doi{10.1051/swsc/2014024}.
\bibitem[{Jakowski et~al.(2012a)Jakowski, B\'eniguel, De~Franceschi, Pajares,
  Jacobsen, Stanislawska, Tomasik, Warnant and Wautelet}]{Jakowski_2012b}
\bibinfo{author}{Jakowski, N.}, \bibinfo{author}{B\'eniguel, Y.},
  \bibinfo{author}{De~Franceschi, G.}, \bibinfo{author}{Pajares, M.},
  \bibinfo{author}{Jacobsen, K.}, \bibinfo{author}{Stanislawska, I.},
  \bibinfo{author}{Tomasik, L.}, \bibinfo{author}{Warnant, R.},
  \bibinfo{author}{Wautelet, G.}, \bibinfo{year}{2012}a.
\newblock \bibinfo{title}{{Monitoring, tracking and forecasting ionospheric
  perturbations using GNSS techniques}}.
\newblock \bibinfo{journal}{J. Space Weather Space Clim.} \bibinfo{volume}{2},
  \bibinfo{pages}{A22}.
\newblock \DOIprefix\doi{10.1051/swsc/2012022}.
\bibitem[{Jakowski et~al.(2012b)Jakowski, Borries and Wilken}]{Jakowski_2012}
\bibinfo{author}{Jakowski, N.}, \bibinfo{author}{Borries, C.},
  \bibinfo{author}{Wilken, V.}, \bibinfo{year}{2012}b.
\newblock \bibinfo{title}{Introducing a disturbance ionosphere index}.
\newblock \bibinfo{journal}{Radio Science} \bibinfo{volume}{47}.
\newblock \DOIprefix\doi{10.1029/2011RS004939}.
\bibitem[{Jakowski and Hoque(2019)}]{Jakowski_2019}
\bibinfo{author}{Jakowski, N.}, \bibinfo{author}{Hoque, M.M.},
  \bibinfo{year}{2019}.
\newblock \bibinfo{title}{Estimation of spatial gradients and temporal
  variations of the total electron content using ground-based gnss
  measurements}.
\newblock \bibinfo{journal}{Space Weather} \bibinfo{volume}{17},
  \bibinfo{pages}{339--356}.
\newblock \DOIprefix\doi{10.1029/2018SW002119}.
\bibitem[{Juan et~al.(2018)Juan, Sanz, Rovira-Garcia, Gonz\'alez-Casado,
  Ib\'a\~nez and Perez}]{Juan_2018}
\bibinfo{author}{Juan, J.}, \bibinfo{author}{Sanz, J.},
  \bibinfo{author}{Rovira-Garcia, A.}, \bibinfo{author}{Gonz\'alez-Casado, G.},
  \bibinfo{author}{Ib\'a\~nez, D.}, \bibinfo{author}{Perez, R.O.},
  \bibinfo{year}{2018}.
\newblock \bibinfo{title}{Aatr an ionospheric activity indicator specifically
  based on gnss measurements}.
\newblock \bibinfo{journal}{J. Space Weather Space Clim.} \bibinfo{volume}{8},
  \bibinfo{pages}{A14}.
\newblock \DOIprefix\doi{10.1051/swsc/2017044}.
\bibitem[{Kakinami et~al.(2013)Kakinami, Yamamoto, Chen, Watanabe, Lin, Liu and
  Habu}]{Kakinami_2013}
\bibinfo{author}{Kakinami, Y.}, \bibinfo{author}{Yamamoto, M.},
  \bibinfo{author}{Chen, C.H.}, \bibinfo{author}{Watanabe, S.},
  \bibinfo{author}{Lin, C.}, \bibinfo{author}{Liu, J.Y.},
  \bibinfo{author}{Habu, H.}, \bibinfo{year}{2013}.
\newblock \bibinfo{title}{Ionospheric disturbances induced by a missile
  launched from north korea on 12 december 2012}.
\newblock \bibinfo{journal}{Journal of Geophysical Research: Space Physics}
  \bibinfo{volume}{118}, \bibinfo{pages}{5184--5189}.
\newblock \DOIprefix\doi{10.1002/jgra.50508}.
\bibitem[{Liu et~al.(2015)Liu, Hu, Wang, Yang, Zhu, Liu, Luhmann and
  Richardson}]{Liu_2015}
\bibinfo{author}{Liu, Y.}, \bibinfo{author}{Hu, H.}, \bibinfo{author}{Wang,
  R.}, \bibinfo{author}{Yang, Z.}, \bibinfo{author}{Zhu, B.},
  \bibinfo{author}{Liu, Y.}, \bibinfo{author}{Luhmann, J.},
  \bibinfo{author}{Richardson, J.}, \bibinfo{year}{2015}.
\newblock \bibinfo{title}{Plasma and magnetic field characteristics of solar
  coronal mass ejections in relation to geomagnetic storm intensity and
  variability}.
\newblock \bibinfo{journal}{The Astrophysical Journal} \bibinfo{volume}{809},
  \bibinfo{pages}{L34}.
\newblock \DOIprefix\doi{10.1088/2041-8205/809/2/l34}.
\bibitem[{Mannucci et~al.(1998)Mannucci, Wilson, Yuan, Ho, Lindqwister and
  Runge}]{Mannucci_1997}
\bibinfo{author}{Mannucci, A.J.}, \bibinfo{author}{Wilson, B.D.},
  \bibinfo{author}{Yuan, D.N.}, \bibinfo{author}{Ho, C.H.},
  \bibinfo{author}{Lindqwister, U.J.}, \bibinfo{author}{Runge, T.F.},
  \bibinfo{year}{1998}.
\newblock \bibinfo{title}{{A global mapping technique for GPS-derived
  ionospheric total electron content measurements}}.
\newblock \bibinfo{journal}{Radio Science} \bibinfo{volume}{33},
  \bibinfo{pages}{565--582}.
\newblock \DOIprefix\doi{10.1029/97RS02707}.
\bibitem[{Mendillo et~al.(2000)Mendillo, Lin and Aarons}]{Mendillo_2000}
\bibinfo{author}{Mendillo, M.}, \bibinfo{author}{Lin, B.},
  \bibinfo{author}{Aarons, J.}, \bibinfo{year}{2000}.
\newblock \bibinfo{title}{The application of gps observations to equatorial
  aeronomy}.
\newblock \bibinfo{journal}{Radio Science} \bibinfo{volume}{35},
  \bibinfo{pages}{885--904}.
\newblock \DOIprefix\doi{10.1029/1999RS002208}.
\bibitem[{Nesterov et~al.(2017)Nesterov, Andreeva, Padokhin, Tumanova and
  Nazarenko}]{Nesterov2017}
\bibinfo{author}{Nesterov, I.A.}, \bibinfo{author}{Andreeva, E.S.},
  \bibinfo{author}{Padokhin, A.M.}, \bibinfo{author}{Tumanova, Y.S.},
  \bibinfo{author}{Nazarenko, M.O.}, \bibinfo{year}{2017}.
\newblock \bibinfo{title}{Ionospheric perturbation indices based on the low-
  and high-orbiting satellite radio tomography data}.
\newblock \bibinfo{journal}{GPS Solutions} \bibinfo{volume}{21},
  \bibinfo{pages}{1679--1694}.
\newblock \DOIprefix\doi{10.1007/s10291-017-0646-1}.
\bibitem[{OMNIWeb(2019)}]{OMNIWeb_data}
\bibinfo{author}{OMNIWeb}, \bibinfo{year}{2019}.
\newblock \bibinfo{title}{Omniweb database}.
\newblock \URLprefix \url{https://omniweb.gsfc.nasa.gov/form/dx1.html}.
\bibitem[{Perevalova et~al.(2016)Perevalova, Edemsky, Timofeeva, Katashevtseva
  and Yasyukevich}]{Perevalova_2016}
\bibinfo{author}{Perevalova, N.}, \bibinfo{author}{Edemsky, I.},
  \bibinfo{author}{Timofeeva, O.}, \bibinfo{author}{Katashevtseva, D.},
  \bibinfo{author}{Yasyukevich, A.}, \bibinfo{year}{2016}.
\newblock \bibinfo{title}{Dynamics of disturbance level of total electron
  content at high and middle latitudes according to gps data}.
\newblock \bibinfo{journal}{Solnechno-Zemnaya Fizika} \bibinfo{volume}{2},
  \bibinfo{pages}{36--43}.
\bibitem[{Perevalova et~al.(2014)Perevalova, Sankov, Astafyeva and
  Zhupityaeva}]{PEREVALOVA_2014}
\bibinfo{author}{Perevalova, N.}, \bibinfo{author}{Sankov, V.},
  \bibinfo{author}{Astafyeva, E.}, \bibinfo{author}{Zhupityaeva, A.},
  \bibinfo{year}{2014}.
\newblock \bibinfo{title}{Threshold magnitude for ionospheric tec response to
  earthquakes}.
\newblock \bibinfo{journal}{Journal of Atmospheric and Solar-Terrestrial
  Physics} \bibinfo{volume}{108}, \bibinfo{pages}{77 -- 90}.
\newblock \DOIprefix\doi{https://doi.org/10.1016/j.jastp.2013.12.014}.
\bibitem[{Perevalova et~al.(2008)Perevalova, Afraimovich, Voeykov and
  Zhivetiev}]{Perevalova_2008}
\bibinfo{author}{Perevalova, N.P.}, \bibinfo{author}{Afraimovich, E.L.},
  \bibinfo{author}{Voeykov, S.V.}, \bibinfo{author}{Zhivetiev, I.V.},
  \bibinfo{year}{2008}.
\newblock \bibinfo{title}{Parameters of large-scale tec disturbances during the
  strong magnetic storm on 29 october 2003}.
\newblock \bibinfo{journal}{Journal of Geophysical Research: Space Physics}
  \bibinfo{volume}{113}.
\newblock \DOIprefix\doi{10.1029/2008JA013137}.
\bibitem[{Pi et~al.(1997)Pi, Mannucci, Lindqwister and Ho}]{Pi_1997}
\bibinfo{author}{Pi, X.}, \bibinfo{author}{Mannucci, A.J.},
  \bibinfo{author}{Lindqwister, U.J.}, \bibinfo{author}{Ho, C.M.},
  \bibinfo{year}{1997}.
\newblock \bibinfo{title}{Monitoring of global ionospheric irregularities using
  the worldwide gps network}.
\newblock \bibinfo{journal}{Geophysical Research Letters} \bibinfo{volume}{24},
  \bibinfo{pages}{2283--2286}.
\newblock \DOIprefix\doi{10.1029/97GL02273}.
\bibitem[{Polyakova and Perevalova(2013)}]{Polyakova_2013}
\bibinfo{author}{Polyakova, A.}, \bibinfo{author}{Perevalova, N.},
  \bibinfo{year}{2013}.
\newblock \bibinfo{title}{Comparative analysis of tec disturbances over
  tropical cyclone zones in the north–west pacific ocean}.
\newblock \bibinfo{journal}{Advances in Space Research} \bibinfo{volume}{52},
  \bibinfo{pages}{1416 -- 1426}.
\newblock \DOIprefix\doi{https://doi.org/10.1016/j.asr.2013.07.029}.
\bibitem[{Sanz et~al.(2014)Sanz, Juan~Z., C.G., C.R., Schlueter, P.R.
  et~al.}]{sanz_2014}
\bibinfo{author}{Sanz, S.}, \bibinfo{author}{Juan~Z., J.M.},
  \bibinfo{author}{C.G., G.}, \bibinfo{author}{C.R., P.},
  \bibinfo{author}{Schlueter, S.}, \bibinfo{author}{P.R., O.}, et~al.,
  \bibinfo{year}{2014}.
\newblock \bibinfo{title}{{Novel ionospheric activity indicator specifically
  tailored for GNSS users}}, \bibinfo{publisher}{The Institute of Navigation,
  Proceedings of ION GNSS+. Tampa, Florida (USA)}. pp.
  \bibinfo{pages}{1173--1182}.
\bibitem[{Stanislawska and Gulyaeva(2015)}]{Stanislawska_2015}
\bibinfo{author}{Stanislawska, I.}, \bibinfo{author}{Gulyaeva, T.},
  \bibinfo{year}{2015}.
\newblock \bibinfo{title}{Ionospheric W Index Based on GNSS TEC in the
  Operational Use for Navigation Systems}.
\newblock \DOIprefix\doi{10.5772/59902}.
\bibitem[{Stankov et~al.(2006)Stankov, Jakowski, Tsybulya and
  Wilken}]{Stankov_2006}
\bibinfo{author}{Stankov, S.M.}, \bibinfo{author}{Jakowski, N.},
  \bibinfo{author}{Tsybulya, K.}, \bibinfo{author}{Wilken, V.},
  \bibinfo{year}{2006}.
\newblock \bibinfo{title}{{Monitoring the generation and propagation of
  ionospheric disturbances and effects on Global Navigation Satellite System
  positioning}}.
\newblock \bibinfo{journal}{Radio Science} \bibinfo{volume}{41}.
\newblock \DOIprefix\doi{10.1029/2005RS003327}.
\bibitem[{Voeykov and Berngardt(2019)}]{Berngardt_2019}
\bibinfo{author}{Voeykov, S.}, \bibinfo{author}{Berngardt, O.},
  \bibinfo{year}{2019}.
\newblock \bibinfo{title}{berng/wtec: Wtec v.1.0.1}.
\newblock \URLprefix \url{https://zenodo.org/record/2604418},
  \DOIprefix\doi{10.5281/ZENODO.2604418}.
\bibitem[{Voeykov et~al.(2016)Voeykov, Berngardt and Shestakov}]{Voeykov_2016}
\bibinfo{author}{Voeykov, S.V.}, \bibinfo{author}{Berngardt, O.I.},
  \bibinfo{author}{Shestakov, N.V.}, \bibinfo{year}{2016}.
\newblock \bibinfo{title}{Use of the index of tec vertical variation
  disturbance in studying ionospheric effects of the chelyabinsk meteorite}.
\newblock \bibinfo{journal}{Geomagnetism and Aeronomy} \bibinfo{volume}{56},
  \bibinfo{pages}{219--228}.
\newblock \DOIprefix\doi{10.1134/S0016793216020122}.
\bibitem[{Wanninger(1993)}]{Wanninger_1993}
\bibinfo{author}{Wanninger, L.}, \bibinfo{year}{1993}.
\newblock \bibinfo{title}{Ionospheric monitoring using igs data},
  \bibinfo{publisher}{Proceedings of the 1993 IGS Workshop Berne, Switzerland:
  Astronomical Institute, Univ. of Berne}. pp. \bibinfo{pages}{351--360}.
\bibitem[{WDC(2019)}]{Kyoto_WDC_data}
\bibinfo{author}{WDC, K.}, \bibinfo{year}{2019}.
\newblock \bibinfo{title}{Kyoto wdc database}.
\newblock \URLprefix \url{http://wdc.kugi.kyoto-u.ac.jp/}.
\bibitem[{Wilken et~al.(2018)Wilken, Kriegel, Jakowski and
  Berdermann}]{Wilken_2018}
\bibinfo{author}{Wilken, V.}, \bibinfo{author}{Kriegel, M.},
  \bibinfo{author}{Jakowski, N.}, \bibinfo{author}{Berdermann, J.},
  \bibinfo{year}{2018}.
\newblock \bibinfo{title}{An ionospheric index suitable for estimating the
  degree of ionospheric perturbations}.
\newblock \bibinfo{journal}{J. Space Weather Space Clim.} \bibinfo{volume}{8},
  \bibinfo{pages}{A19}.
\newblock \DOIprefix\doi{10.1051/swsc/2018008}.
\bibitem[{Zherebtsov and Perevalova(2019)}]{Zherebtsov_2019}
\bibinfo{author}{Zherebtsov, G.A.}, \bibinfo{author}{Perevalova, N.P.},
  \bibinfo{year}{2019}.
\newblock \bibinfo{title}{Gnss potential to monitor unsuccessful spacecraft
  launches}.
\newblock \bibinfo{journal}{GPS Solutions} \bibinfo{volume}{23},
  \bibinfo{pages}{30}.
\newblock \DOIprefix\doi{10.1007/s10291-018-0822-y}.

\end{thebibliography}

\end{document}